# *Windmill droplets*. Optically induced rotation of biphase oil-in-water droplets

Jesús J. del Pozo,[1†] Ana B. Bonhome-Espinosa,[1†] Wei Sun,[1,2] Carlos Gutiérrez-Ariza,[1] Raúl A. Rica,[1,3*] Laura Rodríguez-Arco[1,3,4*]

[1] Universidad de Granada, Department of Applied Physics. Campus de Fuentenueva S/N, 18071, Granada (Spain)

[2] Department of Physics, Yanshan University. 066004, Qinhuangdao (China)

[3] Nanoparticles Trapping Laboratory, Research Unit 'Modeling Nature' (MNat), Universidad de Granada, 18071 Granada, Spain

[4] Instituto de Investigación Biosanitaria Ibs.GRANADA 18014, Granada (Spain)

*e-mail: rul@ugr.es, l_rodriguezarco@ugr.es

[†] These authors contributed equally

## Abstract

In the field of microdroplet manipulation, optical tweezers have been used to form and grow droplets, to transport them, or to measure forces between droplet pairs. However, the exploration of out-of-equilibrium phenomena in optically trapped droplets remains largely uncharted. Here, we report the rotation of biphasic droplets fabricated by co-emulsifying two immiscible liquids (i.e., hydrocarbon and fluorocarbon oils) with a refractive index mismatch in water. When trapped, droplets of a specific geometry rotate around the axis of the laser beam, in what appears to be a dissipative, out-of-equilibrium phenomenon. The rotational frequency, obtained from image analysis is stable and proportional to the beam power. Remarkably, droplets that do not interact with the trapping beam can also be rotated indirectly. This is achieved by positioning the droplets at the center of a circular arrangement of multiple, sequentially activated traps. In this case, the droplet orients towards the location of the active trap by a mechanism yet to be understood. Altogether, our results demonstrate out-from-equilibrium phenomenology in optically trapped biphase droplets, which could inspire the development of devices based on them (e.g., optically induced mixing, etc.). In addition, they may shed light on fundamental principles of optical manipulation of asymmetric particles.

## Introduction

The manipulation of tiny volumes of liquids in the form of droplets is already revolutionizing fields such as biotechnology, healthcare or materials science. The rapid heat and mass transfer within droplets, combined with reduced reagent consumption and waste production, enhances reaction rates and yields, thereby improving sustainability.[1,2] This is particularly advantageous for high-throughput screening of reaction conditions in genomics or drug discovery, and for conducting rapid, on-site testing of minimal sample sizes in diagnostics (e.g., digital PCR or hyperspectral imaging). Additionally, droplet-based templating can be exploited to fabricate complex hierarchical structures, which could find applications in high-performance sensors, energy storage devices, and biomedical implants.[1-6]

In the last decade there has been significant progress in remotely manipulating droplets by external forces (*i.e.*, active manipulation), utilizing electric or magnetic fields, acoustic waves, thermal forces or optical manipulation.[5] These techniques enable control over droplet dynamics such as movement and positioning, fusion and mixing, fission, and sorting. In particular, light-driven methods enable high-resolution spatiotemporal control and reconfigurability, thanks to the ease of tuning the intensity, position shape or pattern of the laser beam. They also eliminate the need of adding electrodes or wiring.[5] Notably, optical tweezers, which utilize a highly focused Gaussian beam to confine, move and rotate particles, stand out in terms of precision, dynamic control, scalability, versatility, and minimal heat.[7,8] In microemulsion droplet technology optical tweezers have been used to form and grow microdroplets,[9,10] to move them to specific locations,[9] or to measure the force between pairs of droplets.[11-13] However, the exploration of out-of-equilibrium phenomena in optically trapped emulsion droplets remains almost unexplored, despite the interest from the fundamental and applied viewpoints of these.[14,15]

In general, when particles are placed near the focus of the gaussian beam in optical tweezers setups, they experience two forces: the gradient force and the scattering force.[16,17] In the case of a dielectric particle whose refractive index ($n_p$) is larger than that of the surrounding medium ($n_m$), the light that impinges the particle from the medium bends (refracts) towards the higher refractive index material, leading to a gradient force that pulls the particle towards the regions of highest intensity, at the focus. The scattering force, caused by the radiation pressure from the momentum transfer of the beam photons, always pushes the particle along the direction of the beam's propagation. Therefore, under appropriate conditions, particles with a refractive index higher than that of the surrounding medium ($n_p > n_m$) experience gradient forces that can overcome scattering forces, trapping the particles at a stable position close to the focus of the beam. Conversely, particles with a lower refractive index than the surrounding medium ($n_p < n_m$) cannot be trapped with this configuration, because the gradient force pushes the particle away from the focus when the refractive index contrast is negative ($n_p - n_m < 0$).[18]

Due to the substantial influence of the refractive index in optical trapping experiments, here we were intrigued by the potential outcome of co-emulsifying two immiscible liquids with refractive index mismatch within a single droplet. These biphase droplets, based on liquid-liquid phase separation, represent a fascinating frontier in droplet technology.[19] The asymmetry intrinsic to these droplets offers unique possibilities in terms of chemical composition and functionalities, enabling each region of the droplet to behave differently in response to external cues. Indeed, asymmetry has been exploited to prompt out-of-equilibrium behaviours, such as droplet self-propulsion.[20,21] Particularly interesting are oil-in-water biphase droplets, where the inner phase consists of a mixture of a hydrocarbon oil and a fluorocarbon oil. From an optical standpoint these are noteworthy, because the refractive index of the hydrocarbon oil is higher than that of water, while the refractive index of the fluorocarbon oil is lower than the refractive index of water.[22-25] In addition, these droplets undergo dynamic morphological transitions in response to external stimuli. Specifically, they can be reconfigured between double emulsion droplets and Janus droplets.[19] This property has been exploited, for example, to fabricate tuneable compound lenses, which can behave as either diverging or focusing lenses depending on the droplet morphology.[22,23]

Prompted by their fascinating properties, here we prepare oil-in-water biphase droplets using hydrocarbon (dodecane) and fluorocarbon (2-(trifluoromethyl)-3-ethoxydodecafluorohexane) oil mixtures as the inner phase. At room temperature, the two oils are immiscible leading to the formation of distinct phases within the droplets, which creates an asymmetrical refractive index profile. We investigate their trapping capabilities in a commercial optical tweezers setup to find out that out-of-equilibrium droplet rotation, not reported previously, arises. Rotation is dependent on the droplet morphology and can be accelerated by increasing the laser beam power. In addition, we demonstrate that it is also possible to externally induce user-defined droplet rotation by shining a laser in the vicinity of the droplets (*i.e.*, when the droplets are not trapped). Overall, our results show complex, out-of-equilibrium interactions in an optical trap setting, which might provide insights into fundamental principles of optical manipulation of asymmetric particles.

**Methods**

**Materials**

n-Dodecane (99% purity for synthesis, Merck USA, density of 0.750 g cm$^{-3}$) and 2-(trifluoromethyl)-3-ethoxydodecafluorohexane (HFE-7500 3M™ Novec™ 7500 Engineered Fluid, Fluorochem, United Kingdom, density of 1.614 g cm$^{-3}$) were used as hydrocarbon and fluorocarbon oils, respectively. The anionic hydrocarbon surfactant sodium dodecyl sulfate (SDS, 95% purity, Scharlab, Spain) was used as surfactant for the dodecane phase, while the fluorinated surfactant Capstone® FS-30 (Apollo Scientific, United Kingdom) was chosen as stabiliser for the HFE-7500 phase.

**Droplet fabrication**

To fabricate the biphase droplets, we first prepared a stock emulsion by dispersing 100 µL of a 1:1 mixture of dodecane and HFE-7500 into 500 µL of an aqueous solution of Capstone (0.007 wt %) and SDS (0.025 wt %). The dispersion was conducted using a homogenizer (IKA Ultra-Turrax T 10 basic, Germany) set at 30 000 rpm for 30 s. To engineer the desired morphology, 5 µL of the stock emulsion was pipetted into 500 µL of a given surfactant solution in water (composition detailed in the table below) to allow for reconfiguration of the droplet. The morphology was left to equilibrate for 24 h before optical trapping experiments.

| Sample | Composition |
| --- | --- |
| FC-in-HC | 0.035 wt % SDS<br>0.004 wt % Capstone |
| HC-in-FC | 0.014 wt % Capstone |

Monophase control samples of dodecane or HFE-7500 alone were prepared by emulsifying 100 µL of the oil in 500 µL of 0.05% wt solutions of SDS and 0.014% wt solutions of Capstone, respectively. For optical tweezers experiments these samples were diluted by pipetting 5 µL of the stock in 500 µL of the same surfactant solutions.

Optical microscopy images of the as-prepared samples (recorded with an Olympus U-TV0.5XC-3 microscope (Achilles, Japan)) are shown in Fig. S1 and S2.

**Optical trapping experiments**

Optical trapping experiments were conducted in a commercial optical tweezers setup (NanoTracker-II from JPK-Bruker, Germany) equipped with a 1064 nm laser. Observation chambers were fabricated by carefully depositing two fine seams of high viscosity sealing paste (Korasilon paste, Kurt Obermeier GmbH, Germany) on a #1.5 glass slide, which was then covered with a smaller #1.5 coverslip. Around 50 µL of the droplet sample were pipetted in between the two glass layers and the chamber was sealed using the same sealing paste.

In the control oil experiments, we focused on a given monophase droplet and tested whether it could be trapped (*i.e.*, dodecane-based) or not (*i.e.*, HFE-based). Movies were recorded. In the case of the biphase droplets, which were always trapped we tested whether windmill rotation arose. In affirmative cases we varied the laser power in the range 10 – 122 mW and recorded videos of the rotation for three different droplets.

In addition, we recorded droplet tilt when the laser beam was focused on the vicinity of the droplets (*i.e.*, in the absence of trapping). In this regard, we also focused the beam at specific locations within the droplet vicinity. To do this, we arranged twelve traps into a circumference pattern, following the layout of a 12-hour clock. The traps were sequentially activated and the dwell time (*i.e.*, the duration for which the laser remained switched on in a particular trap) was varied from 20 ms to 1 s).

**Image analysis**

Analysis of the droplet rotational trajectories and calculation of the rotational frequencies were performed using *Phyton* and the numerical computing/plotting libraries *NumPy* and *Matplotlib*. The *Imageio* library was used to extract metadata from the images.

The script initially applied a frame-by-frame thresholding procedure to create a binary (1-0) image. Thresholding involved a threshold value, and an approximate droplet radius provided by the user. Pixels with intensities above the threshold were set to white (1), while those below were set to black (0). Using this binary image as a starting point, the script then fitted the droplet contour to an ellipse using the user-provided radius value as an initial guess and identified the midpoint between the foci. The script tracked the position of this midpoint and generated a file containing the *x*, *y* coordinates over time. If, for some reason, the script failed to detect an ellipse for a particular frame, it prompted the user to provide new threshold and radius values.

The rotational frequencies of *x* and *y* were determined by fitting plots of these parameters against time with a sine function for each laser power. Fitting was performed for time intervals of 1, 2, 5, 10, 20 and 30 s (sampling times) to find out that the obtained frequencies did not depend on the sampling time (see Fig .S3). The values plotted in the main text represent the average frequency across all sampling times.

## **Results**

To investigate the optical trapping behaviour of droplets with an asymmetrical refractive index distribution, we prepared biphase oil-in-water (O-in-W) droplets where the oil phase was a mixture of a hydrocarbon (HC) oil, dodecane (refractive index $n_{HC}$ = 1.421), and a fluorocarbon (FC) oil, 2-(trifluoromethyl)-3-ethoxydodecafluorohexane (commercially known as HFE-7500 3M™ Novec™ 7500 Engineered Fluid, refractive index $n_{FC}$ = 1.29).

Because these two oils are immiscible at room temperature, they gave rise to phase separation within the droplets, as shown in Fig. 1 a-d. As previously reported, we were able to prepare droplets with two thermodynamically permitted configurations depending on the interfacial tensions of the HC-W, the FC-W, and the HC-FC interfaces, denoted as $\gamma_{HC}$, $\gamma_{FC}$, and $\gamma_{HC-FC}$, respectively. To modulate these, we used mixtures of solutions of a fluorinated surfactant (Capstone) and of a hydrocarbon surfactant (SDS) as aqueous phase for droplet formation. When Capstone was in higher proportion ($\gamma_{HC} - \gamma_{FC} \geq \gamma_{HC-FC}$), the FC oil completely encapsulated the HC oil, forming HC-in-FC droplets (Fig. 1a-b, Fig. S1). Conversely, when SDS predominated ($\gamma_{FC} - \gamma_{HC} \geq \gamma_{HC-FC}$) the HC oil encapsulated the FC oil, resulting in FC-in-HC droplets (Fig. 1c-d, Fig. S1). Monophase control droplets consisting of either HC oil (in a SDS solution) or FC oil (in a Capstone solution) were also fabricated for comparison (Fig. S2, Supporting Information).

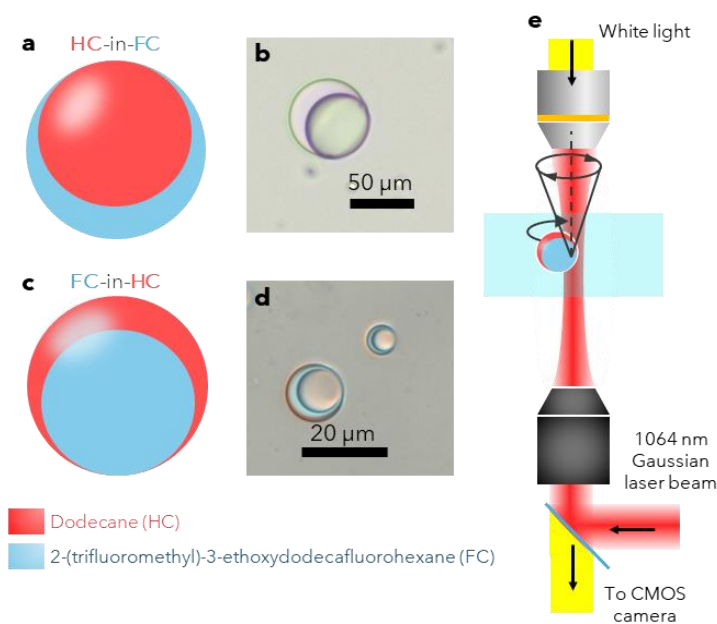

**Figure 1.** Schematics **(a,c)** and optical microscopy images **(b,d)** of oil-in-water biphase droplets with hydrocarbon (HC) and fluorocarbon (FC) oil mixtures as inner phase. Depending on the surfactant composition of the aqueous phase two distinct morphologies could be prepared: droplets in which the FC oil encapsulated the HC oil (HC-in-FC) (**a, b**) or droplets with the FC oil being encapsulated by the HC one (FC-in-HC) (**c, d**). **e.** Schematics of the optical tweezers setup. The droplets were trapped close to the focus of a Gaussian laser beam with wavelength 1064 nm. The droplet was arranged in the trap such that the HC phase was located close to the focus of the beam.

We then inspected the samples in the optical tweezers setup (Fig. 1e) equipped with a steerable laser beam (wavelength λ = 1064 nm, linear polarization). We began by analysing the monophase droplets. HC droplets could be easily trapped close to the beam waist of the laser and dragged along with it (see Fig. 2a and Supplementary Movie S1). We attributed this behaviour to the refractive index mismatch between the dodecane ($n_{HC}$ = 1.421) and water ($n_W$ = 1.33). On the other hand, when FC droplets were placed in the observation chamber optical trapping was not possible. Instead, these droplets were repelled by the laser when the beam was translated across the field of view, resulting in net droplet transport as well (see Fig. 2b and Supplementary Movie S2). This was consistent with the lower refractive index of the FC oil ($n_{FC}$ = 1.29) compared to that of water.[25]

Given the opposite behaviour of the control droplets in their interaction with the laser beam, we wondered how biphase droplets would behave when placed in the optical tweezers setup. Both HC-in-FC and FC-in-HC droplets could be stably trapped with the beam. Remarkably, FC-in-HC droplets started to rotate around the laser beam axis, with stable rotation frequency (see Fig. 2c and Supplementary Movie S3). The motion was neither orbital[26] nor spin[8] like. Instead, we observed that the droplet was trapped at a fixed position, apparently close to the interface between the two oils, and the droplet rotated around this point very much as a Bravais pendulum would do around its suspension point.[27] The rotation was well maintained for long periods (several minutes), although the rotation direction often reversed, *i.e.*, switching from counterclockwise to clockwise (Supplementary Movie S4). Rotation stopped as soon as the laser was switched off, and was accelerated when the power of the beam was increased. On the other hand, HC-in-FC droplets did not show any rotational motion, but enhanced fluctuations which increased with increasing beam power (Supplementary Movie S5).

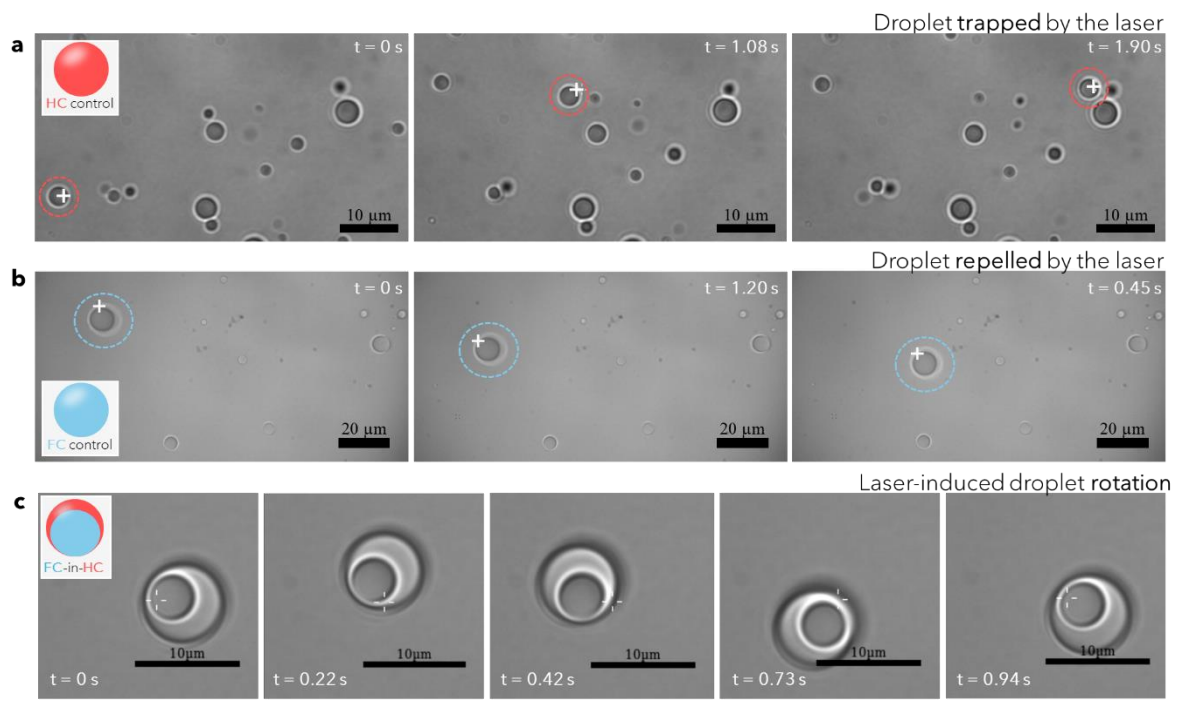

**Figure 2.** Optical microscope snapshots of monophase droplets consisting of pure HC **(a)** and FC **(b)** oils as well as FC-in-HC droplets **(c)** in the optical tweezers chamber; the position of the beam is indicated with a cross symbol. HC oil droplets (refractive index higher than that of water) were successfully trapped and could be moved along with the beam **(a)**. Conversely, FC oil droplets (refractive index lower than that of water) were repelled by the laser, pushing them away from it, which also resulted in net droplet translation **(b)**. On the other hand, FC-in-HC droplets, with an asymmetric distribution of the refractive index, were successfully trapped and experienced an anomalous rotational motion around the axis of the beam of well-defined period (laser power 10 mW) **(c)**.

Encouraged by these results, we analysed the rotation motion of the FC-in-HC droplets by image analysis using a script in Phyton. Briefly, we fitted the droplet contour with an ellipse, identified the middle point between the foci and tracked the position of this point (*i.e.*, its *x*, *y* coordinates). The trajectory plot (*y* vs *x* plot) confirmed that the droplet followed a circular path (Fig. 3a). Both the *x* and *y* coordinates oscillated with well-defined amplitude and period, and approximately constant amplitude (Fig. 3b and c, respectively). We quantified the rotation frequency by fitting the plots of the temporal evolution of *x* and *y* with a sine function for each laser power and confirmed that the frequency was roughly the same for the two parameters (Fig. S3).

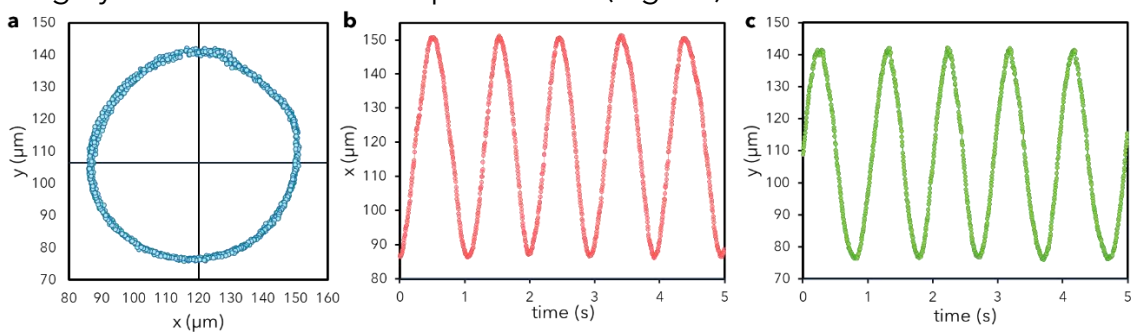

**Figure 3. a.** Trajectory plot (*i.e.*, y *vs* x coordinate plot) of the middle point between the foci of the ellipse to which we fitted the contour of a FC-in-HC rotating droplet using a Python script (see Methods). The data confirm that the droplet experienced a rotational motion at 10 mW of laser power. **b-c.** Temporal evolution of coordinates *x* **(b)** and *y* **(c)** for the same droplet. The two quantities described oscillatory motion characterized by regular amplitude and frequency. All plots correspond to the image analysis of the initial five seconds of droplet rotation.

As described before, when the laser power was increased from 10 to 122 mW, the rotation speed seemed to increase. We corroborated these observations by image analysis and found that the period of oscillation became around four times shorter when the laser power increased from 10 to 37 mW, for example, as shown in Fig. 4a and Supplementary Movies S3 and S4, respectively. On the other hand, the rotation amplitude did not significantly change (Fig. 4a). When we plotted the rotational frequencies of *x* and *y*, that we call $f_x$ and $f_y$, respectively, against the laser power, we found that they followed linear trends. Indeed, when we fitted the data to a straight line, we obtained coefficients of determination $R^2 \geq 0.99$, Fig. 4b. As mentioned before and shown in Fig. 4b the rotational frequencies $f_x$ and $f_y$ were almost coincident, and so they were the values of the linear fitting parameters.

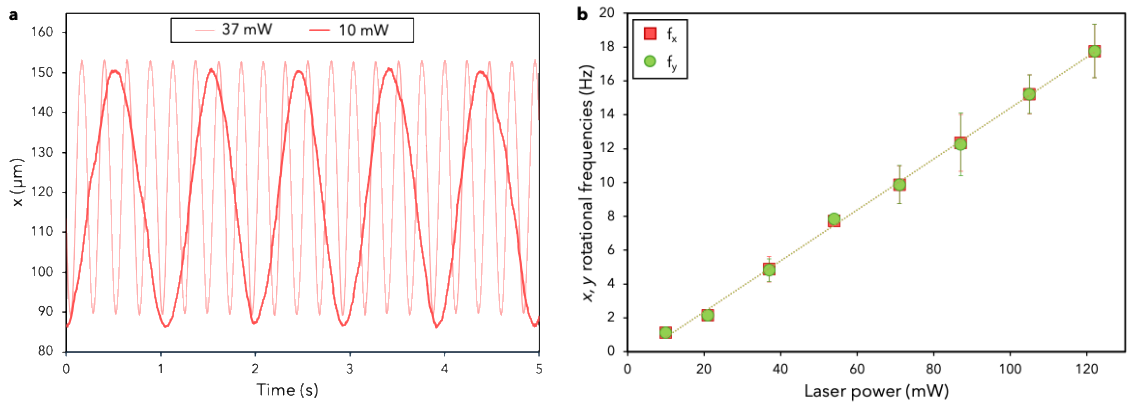

**Figure 4. a.** Plot of the temporal evolution of the *x* coordinate for a FC-in-HC droplet at two different laser powers: 10 and 37 mW. The rotation frequency at 37 mW was roughly four times higher than at 10 mW. **b.** Plots of the rotation frequencies of *x* and *y* against the laser power (P) showing linear trends. Linear fitting equations and determination coefficients are: $f_x$ = (-0.66 ± 0.16) + (0.1506 ± 0.0022)· $P$ ($R^2$ = 0.9985) and $f_y$ = (-0.65 ± 0.18) + (0.150 ± 0.003)· $P$ ($R^2$ = 0.9981). Each data point corresponds to the average frequency obtained by fitting graphs like that in (a) to a sine for 1, 2, 5, 10, 20 and 30 s, for three individual droplets; error bars correspond to standard deviations.

We emphasize that rotation is achieved with a laser beam that is linearly polarized, *i.e.*, not carrying any angular momentum. Such rotation with a linearly polarized beam has been explained in terms of the so-called windmill effect.[28-31] In this situation, the anisotropy of the trapped object is responsible for the deflection of the light rays as they pass through it, and the change in light momentum imparts a torque on the trapped object, initiating the rotation. This torque is counterbalanced by the drag force of the liquid, setting a constant rotation velocity in the steady state. From an experimental point of view, optically induced rotation with light not carrying any angular momentum has only been shown (to the best of our knowledge) in anisotropic micro-objects made with complicated microfabrication techniques (micromachining and two-photon polymerization),[31] and more recently in a triplet of colloidal beads.[32] In the latter, one the beads was thermally active (*i.e.*, absorbing) and generated a temperature gradient which led to imbalanced thermophoretic forces and torques.[32] However, rotation of liquid-based objects has not been demonstrated. The rotation of non-spherical objects like those of our design in optical trapping setups has been theoretically discussed by Pesce *et al.*,[17] providing a brief explanation of a computation method to estimate the forces, which is out of the scope of the present work. Significantly, biphase droplets with the opposite morphology, *i.e.*, HC-in-FC droplets did not rotate when the beam was switched on (see Supplementary Movie S5). We attribute this lack of rotation to the opposite optical characteristics of these droplets compared to the FC-in-HC droplets. Indeed, the FC-in-HC droplets, with a low refractive index core and a higher refractive index shell have been shown to behave as diverging lenses. On the contrary, HC-in-FC droplets behave as convergent, focusing, lenses.[23]

Even though HC-in-FC did not spontaneously rotate in the same out-of-equilibrium, dissipative, way as the FC-in-HC droplets when trapped, both HC-in-FC and FC-in-HC droplets could be externally rotated in a user-defined way if they were not trapped. Indeed, we had observed that when the laser beam was focused in the vicinity of the droplets (*i.e.*, not trying to trap them), both types of droplets tilted with the internal phase of the droplet facing the beam (see Fig. 5a and Supplementary Movie S6). Encouraged by this unexpected behaviour, we set up twelve traps arranged in a circumference like the numbers on a 12-hour clock. When we activated them sequentially with a dwell time of 500 ms, FC-in-HC droplets placed near the centre of the circumference started to rotate following the beam, resembling the hands of the clock (see Fig. 5b and Supplementary Movie S7). Decreasing the dwell time to 50 ms led to faster, rather continuous, rotation (Supplementary Movie S8). HC-in-FC could also be externally rotated in the same way (Supplementary Movie S9). Tilting and motion towards a UV light beam of biphase oil-in-water droplets were recently demonstrated by Frank et al.[20] However, in their case, UV irradiation resulted in non-uniform trans-cis isomerization of the azobenzene-based hydrocarbon surfactant used in droplet formulation. The resulting gradient of hydrocarbon surfactant across the sample led to droplet motion (i.e., Marangoni effect). We did not observe any droplet motion when shinning the laser in the droplet vicinity. Although we hypothesize that droplet tilt might be associated with Marangoni flows as well, future work is needed to elucidate the mechanisms behind this behaviour.[33]

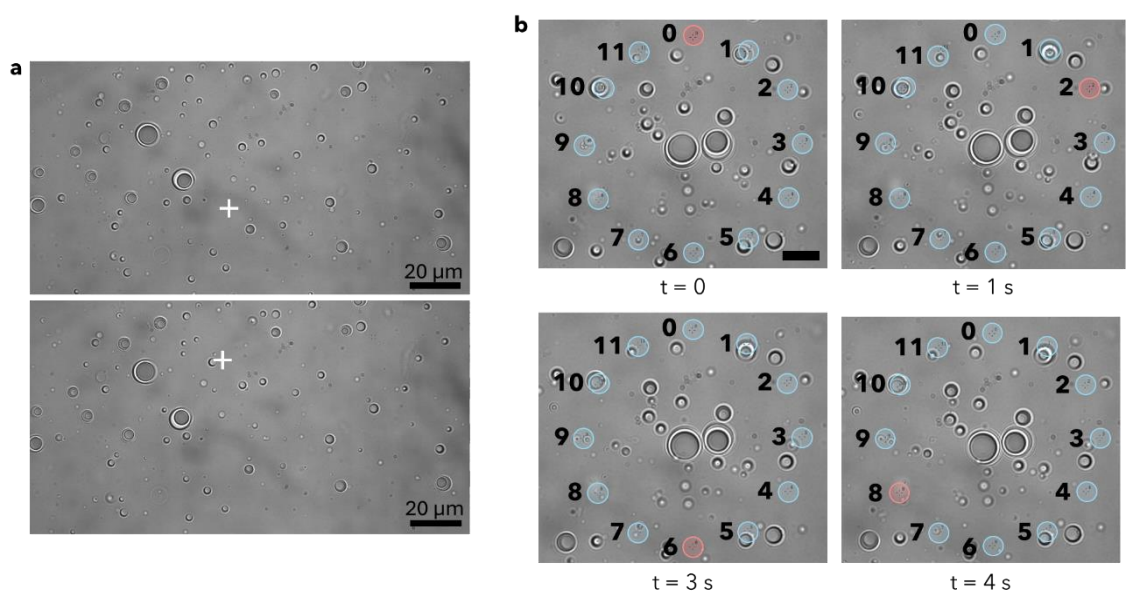

**Figure 5. a.** Optical microscope pictures (from Supplementary Movie S6) showing FC-in-HC droplets tilted with the inner phase facing the laser beam (indicated with a cross symbol). **b.**

Optical microscope snapshots from Supplementary Movie S7 where optical traps (highlighted by blue circles) are arranged into a circumference following a 12-hour clock. The droplets at the centre of the circumference rotate when the traps are sequentially activated (the activated trap is indicated with a red circle) to enable orientation of the inner phase towards the laser beam.

### **Conclusions**

In summary, we have demonstrated an anomalous, out-of-equilibrium, rotation behaviour in biphase droplets placed in an optical trapping setup. These oil-in-water droplets consisted of a mixture of two immiscible oils at room temperature: a hydrocarbon (HC) oil and a fluorocarbon (FC) oil, with refractive indexes higher and lower than the refractive index of water, respectively. We attribute this rotation behaviour to the change in momentum of light that occurs due to the asymmetric distribution of the refractive index within the droplets, that results in a torque applied to the droplet. Notably, only droplets in which the HC oil encapsulated the FC oil (FC-in-HC droplets) rotated with well-defined amplitude and period under the laser beam. The opposite morphology (HC-in-FC droplets) only displayed a slight vibrational motion. We hypothesize that this could be due to the different light interaction of the two morphologies, *i.e.*, FC-in-HC droplets behave as diverging lenses while HC-in-FC droplets focus light.

In addition to out-of-equilibrium rotation when the droplets were trapped, we have shown that droplets that do not interact directly with the trapping beam can also be rotated. This is achieved because when the beam was focused in their vicinity, both types of droplets, HC-in-FC and FC-in-HC, tilted with the inner phase facing the beam. Arranging multiple traps into a circumference and activating them sequentially led to rotation of the droplets. Altogether, our results demonstrate that both out-of-equilibrium and indirect rotation of biphase droplets can be achieved in optical tweezers, which has implications not only from the fundamental point of view, but also in potential applications of the phenomenon, such as optically induced mixing, and related applications that are being developed in the emerging field of optofluidics.[34-37] Future work, including computer simulations, will further elucidate the mechanisms behind the observed phenomenology.

### **Acknowledgements**

The authors acknowledge funding through grant P20_00340 funded by FEDER/Junta de Andalucía - Consejería de Transformación Económica, Industria, Conocimiento y Universidades. A. B. B.-E. also acknowledges co-founding by European Social Fund and Ministry of Economic Transformation, Industry, Knowledge and Universities of the Junta de Andalucía (PAIDI 2020). L.R.-A acknowledges fellowship Juan de la Cierva Incorporación IJC2018-037951-I funded by MCIN/AEI/10.13039/501100011033 and the University of Granada for funding her salary. C.A.G. and R.A.R. acknowledge financial support from Grant PID2021-127427NB-I00 funded by MICIU/AEI/ 10.13039/501100011033 and, by "ERDF A way of making Europe".

## Supplementary information

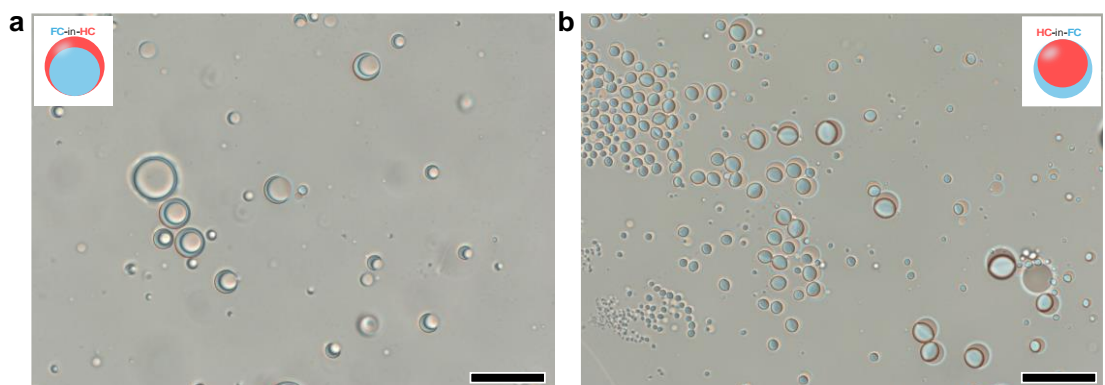

**Figure S1.** Optical microscope pictures of biphase FC-in-HC **(a)** and HC-in-FC **(b)** droplets in aqueous solutions of hydrocarbon (SDS) and fluorinated (Capstone) surfactants. FC-in-HC droplets were obtained in solutions where the hydrocarbon surfactant was in higher proportion, while the opposite morphology was obtained when the fluorinated surfactant was at higher concentration.

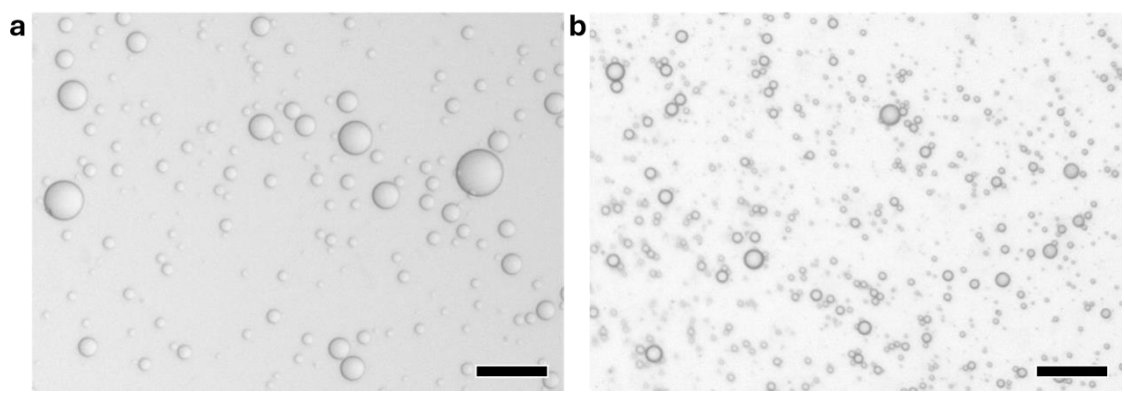

**Figure S2.** Optical microscope pictures of monophase control droplets consisting of either FC oil in an aqueous solution of the fluorocarbon (Capstone) surfactant **(a)** and HC oil in an aqueous solution of the hydrocarbon (SDS) surfactant **(b)**. Scale bars: 50 µm.

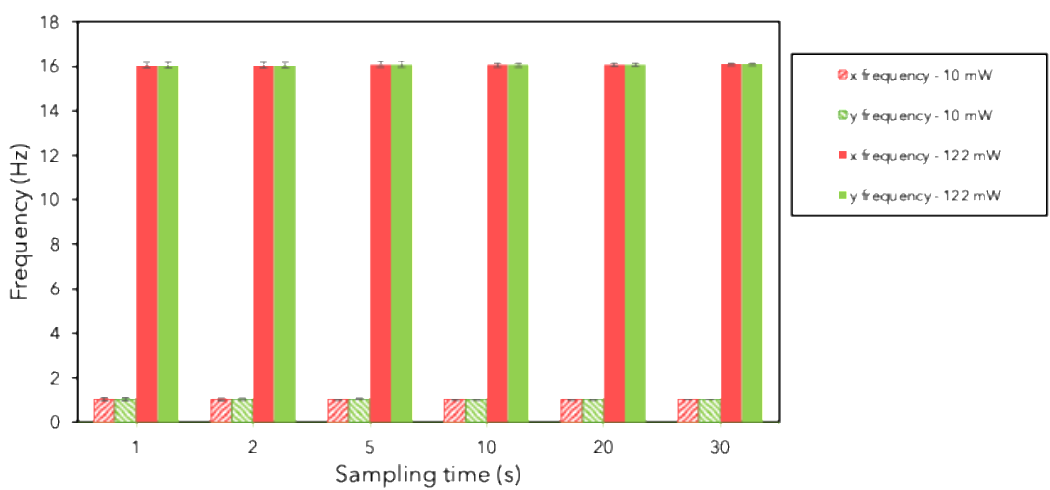

**Figure S3.** Plots of the rotational frequencies of $x$ and $y$ against the sampling time for two different powers of the laser beam (10 mW and 122 mW). The sampling time is the period over which fitting to a sine function of plots like those of Fig. 3 was conducted. As shown, the frequency was almost independent of the sampling time and increased with the laser power.